\documentclass[aps,prb,preprint,superscriptaddress,showpacs,preprintnumbers,amsmath,amssymb,floatfix]{revtex4-1}
\usepackage{graphicx}

\begin{document}

\preprint{}

\title{Dissipative vs dispersive coupling in quantum opto-mechanics: squeezing ability and stability}

\author{A. K. Tagantsev}
\affiliation{ Swiss Federal Institute of Technology (EPFL), CH-1015 Lausanne, Switzerland}
\affiliation{Ioffe Phys.-Tech. Institute, 26 Politekhnicheskaya, 194021, St.-Petersburg, Russia}

\author{I. V. Sokolov}
\affiliation{St Petersburg State Univ, VA Fock Phys Inst, St Petersburg 198504, Russia}

\author{E. S. Polzik}
\affiliation{Niels Bohr Institute, Quantum Optics Laboratory - QUANTOP, Blegdamsvej 17, DK-2100 Copenhagen, Denmark}

\begin{abstract}
Generation of squeezed light and optomechanical instability for dissipative type of opto-mechanical coupling is theoretically addressed for a cavity with the input mirror, serving as a mechanical oscillator, or an equivalent system.
The problem is treated analytically for the case of resonance excitation or small detunings, mainly focusing on the bad cavity limit.
A qualitative difference between the dissipative and purely dispersive coupling is reported.
In particular, it is shown that, for the purely dissipative coupling in the bad cavity regime, the backaction is strongly reduced and the squeezing ability of the system is strongly suppressed, in contrast to the case of purely dispersive coupling.
It is also shown that, for small detunings, stability diagrams for the cases of the purely dispersive and  dissipative couplings are qualitatively identical to within the change of the sign of detuning.
The results obtained are compared with those from the recent theoretical publications.
\end{abstract}
\date{\today}
\maketitle

\section{Introduction}

Cavity quantum optomechanics is a rapidly developing branch of quantum optics which allows for exploration of fundamental issues of quantum mechanics and paves the way for numerous applications, e.g. in high-precision metrology and gravitational-wave defection \cite{Aspelmeyer2014}.
The work horse of cavity optomechanics is the so-called \emph{dispersive coupling} originating from the dependence of the cavity resonance frequency on the position of a mechanical oscillator.
However as pointed out by  Elste et al\cite{Elste2009}, the dispersive coupling does not provide the complete description of the optomechanical interaction.
To fill the gap, those authors have introduced the so-called \emph{dissipative coupling}, which can be interpreted in terms of the dependence of the cavity damping rate on the mirror position.
Since then, manifestations of this coupling have been addressed both theoretically \cite{Xuereb2011,Weiss2013,Weiss2013a,Qu2015,Kilda2016,Tarabrin2013,Vostrosablin2014}
and experimentally \cite{Li2009,Wu2014,Sawadsky2015}.

Studying and harnessing  this effect experimentally seems to be a tough task since it is difficult to find a situation where the dissipative coupling can be distinguished from the stronger dispersive coupling.
Such situation was theoretically identified by Xuereb et al\cite{Xuereb2011} and later experimentally explored by Sawadsky et al\cite{Sawadsky2015} in their setup based on a modified Michelson-Sagnac interferometer where the relative strength of the dispersive and dissipative coupling can be tuned so that the latter can be not dominated by the former.

On the theoretical side, a number of interesting consequences of this coupling have been revealed, e.g. a remarkable Fano effect due to the interference between the dispersive and dissipative couplings \cite{Elste2009,Xuereb2011}.
For the case of dissipative coupling,  detailed zero-temperature numerical simulations of the optomechanical instability as well as of the squeezing and photon correlation spectra have been done by Kilda and Nunnenkamp\cite{Kilda2016}.
One of their results is an interesting possibility of simultaneous squeezing and sideband cooling in the resolved-sideband regime.

It mentioned in the literature \cite{Aspelmeyer2014, Elste2009},  for the dissipative coupling,
 the so-called \emph{bad cavity} regime (the cavity decay rate being larger than the mechanical  resonance frequency) is of special interest, because then this coupling is promising for the ground state cooling of a mechanical oscillator.
A focussed finite-temperature analytical treatment of some manifestations  of the dissipative coupling in this regime was recently published by Qu and Agraval \cite{Qu2015}.
These authors argued that the manifestations of the dissipative coupling they addressed (squeezing spectra and stability conditions) can be quantitatively recovered from the corresponding results for dispersive coupling by replacing the coupling constant of the dispersive coupling with that of the dissipative coupling.
However, some equations and results from Ref.\citenum{Qu2015} are in a conflict with other publications.
Specifically, the stability criterion offered in that paper is incompatible with the results of simulations by Kilda and Nunnenkamp \cite{Kilda2016}.
There is also discrepancy in the Langevin equations and input-output relations between Ref.\citenum{Qu2015}  and  Refs.\citenum{Elste2009,Xuereb2011}.

In view of the above, it seems reasonable to revisit analytically the squeezing generation and the stability conditions of a system with dissipative optomechanical coupling.
This is the subject of the present paper, where we focus on the bad cavity limit.
We have addressed the problem in terms of the Hamiltonian originally introduced by Elste et al\cite{Elste2009}, which is the one exploited in all theoretical papers on the topic. The Hamiltonian describes a one-sided cavity with the input mirror serving as a mechanical oscillator, as well as equivalent systems, including the Michelson-Sagnac interferometer-based setup \cite{Xuereb2011,Sawadsky2015}.

Our analysis demonstrates that, in the bad cavity regime, the purely dissipative coupling manifests itself qualitatively different from the dispersive one.
Namely, in this regime, the backaction due to the dissipative coupling is strongly reduced, vanishing in the limit $\Delta\omega/\gamma\rightarrow 0$ where $\Delta\omega$ and $\gamma$  are the frequency band of interest around the laser frequency and the cavity decay rate, respectively.
Specifically, in this regime, the dissipative coupling constant is effectively multiplied by a factor of the order $\Delta\omega/\gamma$. This effect reveals an additional weakness of the optomechanical interaction in this regime.

Our analytical calculations also provide an explanation for the qualitative difference between the  optomechanical stability diagrams for the purely dispersive and purely dissipative coupling cases \cite{Kilda2016}.
None of our results supports those published by Qu and Agraval \cite{Qu2015}.

The paper is organized as follows.
First, we address, in the simplest terms, the aforementioned reduction of the oscillator-cavity coupling for dissipative case in the bad cavity limit.
Then, we address the implications of this reduction for the squeezing ability of the optomechanical cavity, controlled solely by the dissipative coupling.
We finish the paper with the stability analysis of the system.
Throughout the presentation we illustrate how the behavior of the system controlled by purely dissipative coupling can be mapped onto the system controlled by purely dispersive coupling and then utilize the well known results for the latter system.
\section{Reduction of backaction force in one-sided cavity}
\label{Reduction}
 We start from the Hamiltonian originally introduced by Elste et al\cite{Elste2009}, which describes a one-sided cavity with the input mirror, serving as a mechanical oscillator, and equivalent systems
\begin{equation}
\label{Ham}
\textbf{H}=\hbar\omega_{\textrm{c}}\textbf{a}^{\textbf{+}}\textbf{a}+
\hbar\omega_{\textrm{m}}\textbf{b}^{\textbf{+}}\textbf{b}-\hbar g_\omega\textbf{a}^{\textbf{+}}\textbf{a}(\textbf{b}^{\textbf{+}}+\textbf{b})-
i\hbar\sqrt{\frac{\gamma}{2\pi\rho}} \sum\nolimits_{q}(\textbf{a}^{\textbf{+}}\textbf{c}_q-\textbf{c}_q^\textbf{+}\textbf{a})[1-g_\gamma(\textbf{b}^{\textbf{+}}+\textbf{b})/\gamma],
\end{equation}
where $\hbar$  is the Planck constant, $\omega_{c}$, $\gamma$, and $\textbf{a}$ are the resonance frequency, decay  rate, and the ladder Bose operator of the cavity field, respectively, while  $\omega_{\textrm{m}}$ and $\textbf{b}$ are the resonance frequency and the ladder Bose operator for the mechanical oscillator.
$\textbf{c}_q$ are ladder operators for the electromagnetic bath (the bath Hamiltonian is omitted), $\rho$ is its density of states in the frequency range of interest [per unit frequency].
Here $g_\omega$ and $g_\gamma$ are the coupling constants for the  dispersive and  dissipative interactions, respectively.
The mechanical oscillator is assumed to be coupled to a thermal bath (the Hamiltonian containing the degrees of freedom of the thermal bath is also omitted).

The cavity is driven with a laser light of the frequency $\omega_L$.
Assuming the Markovian bath, one derives in a standard way the Langevin equations describing both the dispersive and dissipative couplings, which (in the frame rotating with the laser frequency) read
\begin{equation}
\label{afull}
\frac{\partial \textbf{a}}{\partial t}+\{\gamma/2+i[\omega_c-\omega_L]\}\textbf{a}
=\sqrt{\gamma}\textbf{A}_{\textrm{in}}+(\textbf{b}^{\textbf{+}}+\textbf{b})\left [ g_\gamma (\textbf{a}-\textbf{A}_{\textrm{in}}/\sqrt{\gamma})+ig_\omega\textbf{a} \right]
\end{equation}
\begin{equation}
\label{bfull}
\frac{\partial \textbf{b}}{\partial t}+\left(\frac{\gamma_{\textrm{m}}}{2}+i\omega_{\textrm{m}}\right)\textbf{b}=\sqrt{\gamma_{\textrm{m}}}\textbf{b}_{\textrm{in}}-
 \frac{g_\gamma}{\sqrt{\gamma}} (\textbf{a}^{\textbf{+}}\textbf{A}_{\textrm{in}}-\textbf{A}_{\textrm{in}}^\textbf{+}\textbf{a})+ig_\omega\textbf{a}^{\textbf{+}}\textbf{a}.
\end{equation}
where $\textbf{A}_{\textrm{in}}$ is the operator of the input electromagnetic field and  $\textbf{b}_{\textrm{in}}$ is the operator of the thermal mechanical noise.
We assume the driving laser field to be strong so that the operators $\textbf{a}$ , $\textbf{b}$ , and $\textbf{A}_{\textrm{in}}$ consist of  constant [in the rotating reference frame] classical parts, $a_0$, $b_0$, $A_0$, and  operator parts, $\delta\textbf{a}$ , $\delta\textbf{b}$ , and $\delta\textbf{A}_{\textrm{in}}$, describing fluctuations.

We are particularly interested in the fluctuations for the case of purely dissipative coupling, i.e. at $g_\omega=0$ and $g_\gamma\neq0$.
The equations describing the fluctuations  are routinely obtained by linearization of Eqs. (\ref{afull}) and (\ref{bfull}).
In the notations where $\delta's$ are dropped from $\delta\textbf{a}$ , $\delta\textbf{b}$, and $\delta\textbf{A}_{\textrm{in}}$ for simplicity, these equations read \cite{Elste2009}:
\begin{equation}
\label{alin}
\frac{\partial \textbf{a}}{\partial t}+\{\gamma/2-i\Delta\}\textbf{a}
=\sqrt{\gamma}\textbf{A}_{\textrm{in}}+g_\gamma (a_0-A_0/\sqrt{\gamma})(\textbf{b}^{\textbf{+}}+\textbf{b})
\end{equation}
\begin{equation}
\label{blin}
\frac{\partial \textbf{b}}{\partial t}+\left(\frac{\gamma_{\textrm{m}}}{2}+i\omega_{\textrm{m}}\right)\textbf{b}=\sqrt{\gamma_{\textrm{m}}}\textbf{b}_{\textrm{in}}+\textbf{F}_{\textbf{\textrm{diss}}}
\end{equation}
where $\Delta=\omega_L -\omega_c$ and the operator of the backaction force (to within a factor of some physical dimension) has a form (c.f. Eq.(5) from Ref.\citenum{Elste2009}):
\begin{equation}
\label{F}
\textbf{F}_{\textbf{\textrm{diss}}}= -\frac{g_\gamma}{\sqrt{\gamma}} (a_0^*\textbf{A}_{\textrm{in}}+ A_0\textbf{a}^\textbf{+}-\textbf{A}_{\textrm{in}}^\textbf{+}a_0-A_0^*\textbf{a}).
\end{equation}
Here $\textbf{A}_{\textrm{in}}$ describes the vacuum noise
\begin{equation}
\label{Aa}
[\textbf{A}(t),\textbf{A}^+(t')]=\delta(t-t') \qquad [\textbf{A}(t),\textbf{A}(t')]=0\qquad <\textbf{A}^+(t)\textbf{A}^+(t')> =0,
\end{equation}
$<...>$ and $[...,...]$ denoting the ensemble averaging and the commutator, respectively, the thermal noise operators obey the same commutation relations while
\begin{equation}
\label{b+b}
\langle\textbf{b}_{\textrm{in}}^+(t)\textbf{b}_{\textrm{in}}(t')\rangle = n_{\textrm{th}}\delta(t-t')
\end{equation}
where $n_{\textrm{th}}=[1 -\exp(\hbar\omega_{\textrm{m}}/T)]^{-1}$, $T$ being the temperature in energy units.

The set of equations (\ref{alin})-(\ref{F}) should be appended with the steady state equation for $a_0$:
\begin{equation}
\label{a0}
\{\gamma/2-i\Delta\}a_0
=\sqrt{\gamma}A_0.
\end{equation}

Now we would like to demonstrate, in the simplest situation, an important result of this paper.
Namely, that the backaction due to the dissipative coupling is strongly suppressed in the bad cavity regime.
For simplicity, we consider the lowest approximation in optomechanical coupling constants, though, as will be shown in Subsect.\ref{Diss}, the same conclusion holds for an arbitrary strength of optomechanical couplings.

Let us evaluate the backaction force in the case where the time derivative and the detuning can be neglected in Eqs. (\ref{alin}) and (\ref{a0}), the requirement, equivalent to the bad cavity limit.
By setting $a_0$ real and taking into account that $A_0\approx\sqrt{\gamma} a_0/2$ we find
\begin{equation}
\label{F1}
\textbf{F}_{\textbf{\textrm{diss}}}\approx  -\frac{g_\gamma}{\sqrt{\gamma}}a_0 [\textbf{A}_{\textrm{in}}-\textbf{A}_{\textrm{in}}^\textbf{+}-(\textbf{a}-\textbf{a}^\textbf{+})\sqrt{\gamma}/2].
\end{equation}
Now taking into account that, within our approximations, Eq.(\ref{alin}) yields
\begin{equation}
\label{Aa0}
\textbf{A}_{\textrm{in}}-\textbf{A}_{\textrm{in}}^\textbf{+}\approx (\textbf{a}-\textbf{a}^\textbf{+})\sqrt{\gamma}/2,
\end{equation}
we see that Eq.(\ref{F1}) implies a zero backaction force in the bad cavity limit.
This is in a sharp contrast with the backaction force  due to the dispersive coupling, which, in the lowest order in the coupling constant, according to (\ref{bfull}) reads:
\begin{equation}
\label{F2}
\textbf{F}_{\textbf{\textrm{disp}}}=ig_\omega a_0 (\textbf{a}+\textbf{a}^\textbf{+})\approx i\frac{2g_\omega a_0}{\sqrt{\gamma}} (\textbf{A}+\textbf{A}^\textbf{+}).
\end{equation}

One can readily trace the origin of this difference.
In the case of dissipative coupling, the vacuum electromagnetic noise reaches the mechanical oscillator via two channels: from the cavity field and directly (see Eq.(\ref{F1})).
In the bad cavity regime, a distractive interference between those channels takes place, resulting in the backaction force reduction.
At the same time, in the case of dispersive coupling, nothing similar can happen because now  the vacuum electromagnetic noise reaches the mechanical oscillator via the only channel - the cavity field (see Eq.(\ref{F2})).
Evidently, the above effect entails a reduction of squeezing ability of the dissipative coupling in the bad cavity limit since it is the backaction that provides mixing of the quadratures of the output light, leading to squeezing of the optimized quadrature.
We will address this issue in detail in Subsect.\ref{Diss} below.
\section{Squeezing}
In this section, we compare  the squeezing abilities of the dispersive and dissipative couplings for a single-ended cavity with a mechanical oscillator as the input mirror (or an equivalent system) in the bad cavity limit.
To make presentation more transparent we consider the case of a resonance excitation of the cavity.
In Subsect.\ref{Disp}  we reproduce the well known result for the dispersive coupling as a benchmark.
In Subsect.\ref{Diss} we map the squeezing problem for the case of the dissipative coupling onto that for the dispersive one and,  without further calculations, we recover  the results for the dissipative coupling from those obtained in Subsect.\ref{Disp}.
\subsection{Dispersive coupling}
\label{Disp}
To obtain the squeezing spectrum of the generalized quadrature of the light  backscattered from the cavity, we perform the Fourier transforms on all operators describing fluctuations and obtain for the frequency components, e.g.
\begin{equation}
\label{Fourier1}
\textbf{a}(\omega)= \frac{1}{\sqrt{2\pi} }\int dte^{i\omega t} \textbf{a}(t)
\end{equation}
 \begin{equation}
\label{Fourier2}
 \textbf{a}^\textbf{+}(\omega)= \frac{1}{\sqrt{2\pi} }\int dte^{-i\omega t} \textbf{a}^\textbf{+}(t).
\end{equation}
We introduce the quadrature operators for all  variables (including the noise operators) according to the following rules
$$
\textbf{Q}(\omega)=[\textbf{b}(\omega)+\textbf{b}^\textbf{+}(-\omega)]/2 \qquad \textbf{P}(\omega)=-i[\textbf{b}(\omega)-\textbf{b}^\textbf{+}(-\omega)]/2
$$
$$
\textbf{Q}_{\textrm{in}}(\omega)=[\textbf{b}_{\textrm{in}}(\omega)+\textbf{b}_{\textrm{in}}^\textbf{+}(-\omega)]/2
$$
\begin{equation}
\label{XY}
\textbf{X}(\omega)=[\textbf{a}(\omega)+\textbf{a}^\textbf{+}(-\omega)]/2\qquad \textbf{Y}(\omega)=-i[\textbf{a}(\omega)-\textbf{a}^\textbf{+}(-\omega)]/2
\end{equation}
\begin{equation}
\label{XinYin}
\textbf{X}_{\textrm{in}}(\omega)=\textbf{A}_{\textrm{in}}(\omega)+\textbf{A}_{\textrm{in}}^\textbf{+}(-\omega)
\qquad \textbf{Y}_{\textrm{in}}(\omega)=-i[\textbf{A}_{\textrm{in}}(\omega)-\textbf{A}_{\textrm{in}}^\textbf{+}(-\omega)].
\end{equation}
The Fourier components of the noise operators satisfy the relationships:
\begin{equation}
\label{XYvaromega}
\langle\textbf{X}_{\textrm{in}}(\omega)\textbf{X}_{\textrm{in}}(\omega')\rangle=\langle\textbf{Y}_{\textrm{in}}(\omega)\textbf{Y}_{\textrm{in}}(\omega')\rangle
=i\langle\textbf{Y}_{\textrm{in}}(\omega)\textbf{X}_{\textrm{in}}(\omega')\rangle=
\end{equation}
$$
-i\langle\textbf{X}_{\textrm{in}}(\omega)\textbf{Y}_{\textrm{in}}(\omega')\rangle=\delta(\omega+\omega').
$$
\begin{equation}
\label{QinF}
\langle\textbf{Q}_{\textrm{in}}(\omega)\textbf{Q}_{\textrm{in}}(\omega')\rangle=(n_{\textrm{\textrm{th}}}+1/2)\delta(\omega+\omega').
\end{equation}

Starting from  the standard linearized Langevin equations for fluctuations in the optomechanical cavity controlled by the dispersive coupling, e.g. from Ref.\cite{Weiss2013}, we rewrite those in the quadrature variables (rotating reference frame), dropping the frequency argument $\omega$ wherever it is not confusing:
\begin{align}
\begin{aligned}
(\gamma/2-i\omega)\textbf{X}
=\frac{\sqrt{\gamma}}{2}\textbf{X}_{\textrm{in}}
\end{aligned}
\label{Xa1}
\end{align}
\begin{align}
\begin{aligned}
(\gamma/2-i\omega)\textbf{Y}
=\frac{\sqrt{\gamma}}{2}\textbf{Y}_{\textrm{in}}+G_\omega \textbf{Q}
\end{aligned}
\label{Ya1}
\end{align}
\begin{equation}
\label{Qa1}
\chi(\omega)^{-1}\textbf{Q}=
\sqrt{\gamma_{\textrm{m}}}\textbf{Q}_{\textrm{in}}+G_{\omega}\textbf{X}
\end{equation}
\begin{equation}
\label{xi}
\chi(\omega)^{-1}=\frac{1}{\omega_{\textrm{m}}}\left[\omega_{\textrm{m}}^2-\omega^2 -i\gamma_{\textrm{m}}\omega\right].
\end{equation}
\begin{equation}
\label{Gomega}
G_{\omega}=2g_\omega a_0
\end{equation}
where $a_0$ is the dimensionless amplitude of the classical field in the cavity (set to be real).
Here we have also neglected the  small renormalization of the mechanical frequency when passing from $\textbf{b}$ to $\textbf{Q}$ operators.

The squeezing of the backscattered light in the dissipative coupling  regime is evaluated by the variance of the generalized quadrature
\begin{equation}
\label{Z}
\textbf{Z}(\omega,\theta)=\textbf{X}_{\textrm{out}}(\omega)\cos\theta+\textbf{Y}_{\textrm{out}}(\omega)\sin\theta
\end{equation}
where $\textbf{X}_{\textrm{out}}$ and $\textbf{Y}_{\textrm{out}}$  are defined by the standard input-output relations
\begin{equation}
\label{IOX}
\textbf{X}_{\textrm{in}}+\textbf{X}_{\textrm{out}}=2\sqrt{\gamma}\textbf{X}
\end{equation}
\begin{equation}
\label{IOY}
\textbf{Y}_{\textrm{in}}+\textbf{Y}_{\textrm{out}}=2\sqrt{\gamma}\textbf{Y}.
\end{equation}
Taking  the bad cavity limit $\gamma\gg\omega_{\textrm{m}}$ and keeping  in mind that, in the situation of interest, $|\omega|$ and $\omega_{\textrm{m}}$ are of the same order, the  above formulae
readily yield the explicit input-output relations
\begin{equation}
\label{Xout}
\textbf{X}_{\textrm{out}}=\textbf{X}_{\textrm{in}}
\end{equation}
\begin{equation}
\label{Yout}
\textbf{Y}_{\textrm{out}}=\textbf{Y}_{\textrm{in}}+
\frac{4G_{\omega}}{\sqrt{\gamma}}\chi(\omega)\left[\sqrt{\gamma_{\textrm{m}}}\textbf{Q}_{\textrm{in}}+\frac{G_{\omega}}{\sqrt{\gamma}}\textbf{X}_{\textrm{in}}
\right].
\end{equation}
Then, straightforward calculations yield
\begin{equation}
\label{ZZ}
<\textbf{Z}(\omega,\theta)\textbf{Z}(\omega',\theta)>=\delta(\omega+\omega')S_{ZZ}(\omega,\theta)
\end{equation}
\begin{equation}
\label{SZZfinal1}
S_{ZZ}(\omega, \theta)= 1 +M\sin^2\theta + N\sin\theta\cos\theta
\end{equation}
\begin{equation}
\label{M}
M= 16n_{\textrm{\textrm{ba}}}\gamma_{\textrm{m}}^2|\chi(\omega)|^2(n_{\textrm{\textrm{th}}}+n_{\textrm{\textrm{ba}}} +1/2)
\end{equation}
\begin{equation}
\label{N}
N= 8n_{\textrm{\textrm{ba}}}\gamma_{\textrm{m}}\textrm{Re}[\chi(\omega)].
\end{equation}
Here
\begin{equation}
\label{nba}
n_{\textrm{ba}}= \frac{G_{\omega}^2}{\gamma_{\textrm{m}}\gamma}
\end{equation}
is the optomechanical cooperativity  or, alternatively, the noise added by the backaction to the intrinsic noise of the mechanical oscillator (normalized to the number of mechanical quanta).

Since it is the even part of the spectral power density that is used for characterization of squeezing, hereafter, for this variable we will keep only the frequency-even parts.

The result of minimization of the spectral power density  of the generalised quadrature $S_{ZZ}(\omega, \theta)$ with respect to the "mixing" angle $\theta$, $S_{\textrm{m}}(\omega)$, reads
\begin{equation}
\label{SminS}
S_{\textrm{m}}(\omega)= 1 -\frac{N^2/2}{\sqrt{M^2+N^2}+M}.
\end{equation}
The results of further minimization (with respect to $\omega$) can be presented in a transparent form if we focus on the frequency range close to the mechanical frequency.
Specifically for $\delta= \omega_{\textrm{m}} -\omega$ satisfying the following inequalities
\begin{equation}
\label{ineq1}
|\delta|\ll\omega_{\textrm{m}}
\end{equation}
and
\begin{equation}
\label{ineq2}
\frac{|\delta|/\gamma_{\textrm{m}}}{n_{\textrm{\textrm{th}}}+n_{\textrm{\textrm{ba}}} +1/2}<<1.
\end{equation}
These inequalities allow us to neglect the $N^2$-term under the square root in Eq.(\ref{SminS}) and present this equation in the form
\begin{equation}
\label{SminS1}
S_{\textrm{m}}(\omega)=\frac{n_{\textrm{\textrm{th}}} +1/2}{n_{\textrm{\textrm{ba}}}+n_{\textrm{\textrm{th}}} +1/2}+\frac{n_{\textrm{ba}}}{n_{\textrm{\textrm{ba}}}+n_{\textrm{\textrm{th}}}+1/2}
\frac{\textrm{Im}[\chi(\omega)]^2}{|\chi(\omega)|^2}
\end{equation}
Keeping in mind that in the frequency range of interest
$$
\frac{\textrm{Im}[\chi(\omega)]^2}{|\chi(\omega)|^2}\approx\frac{1}{1+(2\delta/\gamma_{\textrm{m}})^2},
$$
Eq. (\ref{SminS}) can be further simplified to get:
\begin{equation}
\label{SminS2}
S_{\textrm{m}}=\frac{n_{\textrm{\textrm{th}}} +1/2}{n_{\textrm{\textrm{ba}}}+n_{\textrm{\textrm{th}}} +1/2}+\frac{n_{\textrm{ba}}}{n_{\textrm{\textrm{ba}}}+n_{\textrm{\textrm{th}}} +1/2}\frac{1}{1+(2\delta/\gamma_{\textrm{m}})^2} .
\end{equation}
This equation implies that, in the potentially quite wide  frequency range defined by the conditions
$$
1 \ll|\delta|/\gamma_{\textrm{m}}\ll n_{\textrm{\textrm{th}}}+n_{\textrm{\textrm{ba}}}, \omega_{\textrm{m}}/\gamma_{\textrm{m}},
$$
the squeezing parameter of the optimized generalized quadrature approaches a limiting value of
\begin{equation}
\label{SminS3}
S_{\textrm{0}}=\frac{n_{\textrm{\textrm{th}}} +1/2}{n_{\textrm{\textrm{ba}}}+n_{\textrm{\textrm{th}}} +1/2}.
\end{equation}.

This result is consistent with the well known result by Fabre et al \cite{Fabre1994}.
\subsection{Dissipative coupling}
\label{Diss}
Now we keep all settings used in  the previous Subsection the same, except we consider the dissipative coupling instead of the dispersive one.
The linearized Langevin equations for this system in terms of the ladder operators can be found in Refs.\citenum{Weiss2013,Elste2009}.
We rewrite these equations in the quadrature variables:
\begin{align}
\begin{aligned}
(\gamma/2-i\omega)\textbf{X}
=\frac{\sqrt{\gamma}}{2}\textbf{X}_{\textrm{in}}+G_\gamma \textbf{Q}
\end{aligned}
\label{Xa}
\end{align}
\begin{align}
\begin{aligned}
(\gamma/2-i\omega)\textbf{Y}
=\frac{\sqrt{\gamma}}{2}\textbf{Y}_{\textrm{in}}
\end{aligned}
\label{Ya}
\end{align}
\begin{equation}
\label{Qa}
\chi(\omega)^{-1}\textbf{Q}=
\sqrt{\gamma_{\textrm{m}}}\textbf{Q}_{\textrm{in}}+G_{\gamma}(\textbf{Y}-\textbf{Y}_{\textrm{in}}/\sqrt{\gamma})
\end{equation}
where
\begin{equation}
\label{Ggamma}
G_{\gamma}=g_\gamma a_0.
\end{equation}
The difference between the structure of the above mechanical equation and that for the dispersive coupling, Eq.(\ref{Qa1}), is clearly seen: in the mechanical equation corresponding to dissipative coupling, we observe a  direct contribution of the vacuum noise, which is responsible for the strong suppression of the backaction in the bad cavity limit, discussed in Sect. \ref{Reduction}.
The input-output relations \cite{Xuereb2011} also differ from those for the dispersive coupling by an explicit  appearance of the mechanical variable:
\begin{equation}
\label{INOUT1}
\textbf{X}_{\textrm{in}}+\textbf{X}_{\textrm{out}}=2\sqrt{\gamma}\textbf{X}-\frac{4}{\sqrt{\gamma}}G_\gamma \textbf{Q}
\qquad
\textbf{Y}_{\textrm{in}}+\textbf{Y}_{\textrm{out}}=2\sqrt{\gamma}\textbf{Y}.
\end{equation}

Based on the above equations, we find the following explicit input-output  relations, keeping only the lowest non-vanishing terms in $\omega/\gamma$:
\begin{equation}
\label{Xout1}
\textbf{Y}_{\textrm{out}}=\textbf{Y}_{\textrm{in}}
\end{equation}
\begin{equation}
\label{Yout1}
\textbf{X}_{\textrm{out}}=\textbf{X}_{\textrm{in}}+
\frac{4G_{\gamma}\beta(\omega)}{\sqrt{\gamma}}\chi(\omega)\left[\sqrt{\gamma_{\textrm{m}}}\textbf{Q}_{\textrm{in}}
+\frac{G_{\gamma}\beta(\omega)}{\sqrt{\gamma}}\textbf{Y}_{\textrm{in}}\right]
\qquad \beta(\omega)=\frac{2i\omega}{\gamma}.
\end{equation}

Relation (\ref{Yout}) is worth  commenting on.
The small factor $\omega/\gamma$ enters this relation two times: inside and outside  the brackets.
Its first appearance describes the backaction reduction discussed in terms of Langevin equations in Sect.\ref{Reduction}.
The second appearance describes the reduced ability of the system to read the position of the oscillator.
This is a result of cancelation due to the presence of the $\textbf{Q}$ operator in the input-output relations (\ref{INOUT1}).
Remarkably, for measurements, we cannot profit from the reduction of the backaction since the informative signal is also reduced.

Comparing this set of equations with the set (\ref{Yout}) and (\ref{Xout}), we note that those sets are equivalent to each other within swapping
$$
\textbf{X}\Leftrightarrow \textbf{Y}\qquad G_{\omega}\Leftrightarrow  G_{\gamma}\beta(\omega).
$$

This correspondence implies that the well known  results reproduced in  the previous Subsection, Eq.(\ref{SminS1}), can be, with a proper modification, applied to the system with the dissipative coupling.
Specifically, we can write
\begin{equation}
\label{SminS41}
S_{\textrm{m}}=\frac{n_{\textrm{\textrm{th}}} +1/2}{n_{\textrm{\textrm{ba1}}}+n_{\textrm{\textrm{th}}} +1/2}+\frac{n_{\textrm{ba1}}}{n_{\textrm{\textrm{ba1}}}+n_{\textrm{\textrm{th}}} +1/2}\frac{\textrm{Im}[\chi(\omega)]^2}{|\chi(\omega)|^2} .
\end{equation}
where
\begin{equation}
\label{nba1}
n_{\textrm{ba1}}= \frac{G_{\gamma}^2}{\gamma_{\textrm{m}}\gamma}\left(\frac{2\omega}{\gamma}\right)^2,
\end{equation}
plays a role of the optomechanical  cooperativity.

There exists one more difference between the dispersive and dissipative system: swapping $\textbf{X}\Leftrightarrow \textbf{Y}$ leads to the permutation $\cos\theta\Leftrightarrow \sin\theta$ in Eq.(\ref{SZZfinal1}).
That means that, after the  substitution $G_{\omega}\Leftrightarrow G_{\gamma}\beta(\omega)$, the expressions for the maximal squeezing are identical, but the expressions for  the optimal angle are not.

Similarly to the previous case, we thus find  that, in the frequency range defined by conditions
$$
1 \ll\delta/\gamma_{\textrm{m}}\ll n_{\textrm{\textrm{th}}}+n_{\textrm{\textrm{ba1}}}, \omega_{\textrm{m}}/\gamma_{\textrm{m}}\qquad \delta=|\omega-\omega_{\textrm{m}}|,
$$
the squeezing parameter of the optimized generalized quadrature approaches a limiting value of
\begin{equation}
\label{Sminm}
S_{\textrm{0}}=\frac{n_{\textrm{\textrm{th}}} +1/2}{n_{\textrm{\textrm{ba1}}}+n_{\textrm{\textrm{th}}} +1/2}.
\end{equation}.

The result for the optomechnical cooperativity of the system, (\ref{nba1}),
in combination with (\ref{SminS41}) suggests that the bad cavity regime is unfavorable for the squeezing ability of the system.

Our results are in a conflict with those obtained by Qu and Agraval \cite{Qu2015} for squeezing in the dissipative system at resonance in the bad cavity limit.
In the notations of our paper, the result by Qu and Agraval misses the important factor of $\beta(\omega)$ in the definition of $n_{\textrm{ba1}}$.
It does not seem feasible to fully clarify the origin of this disparity.
However, one  problem is clearly seen in Ref. \citenum{Qu2015}:  the vital term with the $\textbf{Q}$ operator in the input-output relations (\ref{INOUT1}) is neglected.
\section{Stability}
\begin{figure}
\includegraphics [width=0.9\columnwidth,clip=true, trim=0mm 0mm 0mm 0mm] {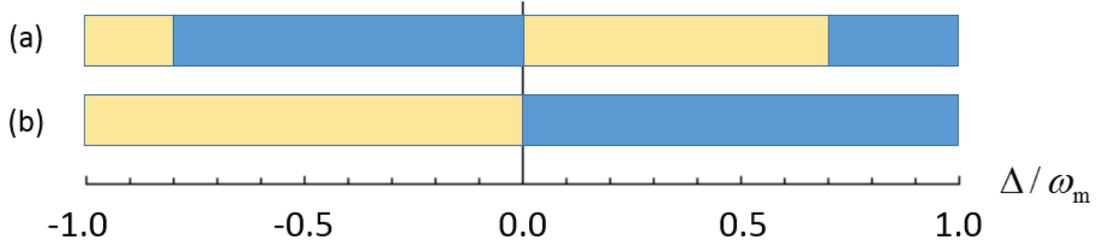}
\caption{Stability diagram for a one-sided cavity with the input mirror, serving as mechanical oscillator, or an equivalent system, according to Kilda and Nunnenkamp \cite{Kilda2016}.
Dark areas show instability intervals in terms of the normalized detuning $\Delta/\omega_\textrm{m}$ ($\Delta =\omega_\textrm{L}-\omega_\textrm{c}$ - detuning, $\omega_\textrm{m}$- mechanical frequency).
(a)- purely dissipative coupling, (b) - purely dispersive coupling.
In notations of our paper, the model settings are $G_\omega=1.2\gamma$, $G_\gamma=-0.3\gamma$, $\gamma_\textrm{m}/\omega_\textrm{m}=10^{-5}$, $\gamma/\omega_\textrm{m}=0.3$; $\gamma$ - decay rate of the cavity, $\gamma_\textrm{m}$ - mechanical decay rate.
\label{f4}}
\end{figure}

For the case of purely dispersive coupling,  the stability of a one-sided cavity with the input mirror serving as the mechanical oscillator has been treated by many authors (see e.g.  Ref. \citenum{Fabre1994}).
At the same time, for the case of purely dissipative coupling, described by Hamiltonian offered by Elste et al \cite{Elste2009}, the stability problem was addressed only recently: numerically by Nunnenkamp with coworkers \cite{Weiss2013,Weiss2013a,Kilda2016} and analytically by Qu and Agraval \cite{Qu2015}, in the bad cavity limit.
According to the numerical calculations\cite{Kilda2016} (see Fig.\ref{f4}a), the system with  purely dissipative coupling is unstable with respect to small red detuning,  in contrast to  the small-blue-detuning instability for the dispersive coupling (see Fig.\ref{f4}b).
In contrast, according to  analytical calculations by Qu and Agraval \cite{Qu2015}, a small blue detuning can destabilize both systems, while a small red detuning is safe.

To resolve this disparity, we show  how, in the limit of small detunings,  the instability problems for those two couplings can be mapped onto each other.

For the dispersive system the instability problem reduces to the analysis of the linear set of equations following, e.g. from Ref.\citenum{Kilda2016}:
\begin{align}
\begin{aligned}
\\&\dot{\textbf{X}}= -\frac{\gamma}{2}\textbf{X}-\Delta\textbf{Y}
\\&\dot{\textbf{Y}}= \Delta\textbf{X}-\frac{\gamma}{2}\textbf{Y}+G_\omega \textbf{Q}
\\&\dot{\textbf{Q}}= \omega_\textrm{m}\textbf{P}
\\&\dot{\textbf{P}}= G_\omega\textbf{X}- \omega_\textrm{m}\textbf{Q}-\gamma_\textrm{m}\textbf{P}.
\end{aligned}
\label{STAB1}
\end{align}
where the dot means the time derivative.
(In this section, the bold-type letters are used for the time-dependent operators, not their Fourier transforms as in the previous sections.)
The application of Routh-Hurwitz criterion \cite{DeJesus1987} for small detuning (only linear terms in $\Delta$ are kept) readily yields the stability  condition:
\begin{equation}
\label{STAB2}
\frac{\Delta}{\omega_\textrm{m}}< \left(\frac{\gamma}{G_\omega}\right)^2 \frac{\gamma}{\omega_\textrm{m}}Q\
\left[1+4Q\left(\frac{\omega_\textrm{m}}{\gamma}\right)^3+16\left(\frac{\omega_\textrm{m}}{\gamma}\right)^4 \right] \qquad Q=\gamma_\textrm{m}/\omega_\textrm{m},
\end{equation}
here we have also neglected $\gamma_\textrm{m}$ compared to other frequency-units parameters.
This relation is consistent with the well known results from  Ref. \citenum{Fabre1994}.
It also complies  well with  the results of modelling by Kilda and Nummenkamp \cite{Kilda2016}, yielding the instability threshold $\frac{\Delta}{\omega_\textrm{m}}=+4*10^{-3}$, c.f. Fig.\ref{f4}b.
From the practical side, this condition means that the system is formally stable for the resonant excitation, however, a small blue detunings jeopardize its stability.

For the dissipative coupling, using the results from Ref.\citenum{Kilda2016}, we find, in the linear approximation in detuning $\Delta$\cite{footnote1}, a similar set of equations for the stability analysis:
\begin{align}
\begin{aligned}
\\ &\dot{\textbf{X}}= -\frac{\gamma}{2}\textbf{X}-\Delta\textbf{Y}+G_\gamma \textbf{Q}
\\&\dot{\textbf{Y}}= \Delta\textbf{X}-\frac{\gamma}{2}\textbf{Y}
\\&\dot{\textbf{Q}}= \omega_\textrm{m}\textbf{P}
\\&\dot{\textbf{P}}= G_\gamma\textbf{Y}- \omega_\textrm{m}\textbf{Q}-\gamma_\textrm{m}\textbf{P}.
\end{aligned}
\label{STAB3}
\end{align}
Comparing the instability problem given by Eq.(\ref{STAB1}) with that given by  Eq.(\ref{STAB3}) we clearly see that those two are equivalent to each other within swapping
$$
G_\omega\Leftrightarrow G_\gamma\qquad \Delta\Leftrightarrow -\Delta.
$$
The transformation $\Delta\Leftrightarrow -\Delta$ explains the complimentarily of the instability regions seen in  Fig.\ref{f4}a and b.
Specifically, it implies that, in contrast to the dispersive-coupling system, in the dissipative-coupling system, small red detunings lead to instability, in contrast to the result obtained by Qu and Agraval \cite{Qu2015}.
%
\section{Discussion and conclusions}

 We have theoretically addressed the squeezing generation and optomechanical instability due to  dissipative coupling for a one-sided optomechanical cavity or an equivalent system.
The Hamiltonian introduced by Elste et al\cite{Elste2009} has been used as the starting point of the analysis.
We have focused on the case of the resonance or small (compared to the decay rate) detuning and the bad cavity regime.
We have identified a remarkable qualitative difference in the manifestations of the purely dissipative and purely dispersive couplings.
We have shown that, in this regime, for the purely dissipative coupling, the backaction is strongly reduced while the squeezing ability of the system is strongly suppressed, in contrast to the case of purely dispersive coupling.
This implies that, for the dissipative coupling system, this regime is extremely unfavorable for the squeezing purposes.
Our results also provide qualitative explanation for the numerical results on the optomechanical instability obtained by Nunnenkamp and coworkers \cite{Weiss2013,Weiss2013a,Kilda2016}.
Specifically, we have demonstrated that, for small detuning, the stability diagrams for the cases of purely dispersive and purely dissipative coupling are complimentary.

Our results for the case of purely dissipative coupling apply to the setup based on a modified Michelson-Sagnac interferometer \cite{Xuereb2011}, where the relative strength of the dispersive and dissipative coupling can be tuned so that the purely dissipative-coupling regime becomes experimentally feasible \cite{Sawadsky2015}.


\bibliography{QOwork,NF}

\begin{thebibliography}{15}
\bibitem [{foo()}]{footnote1}%
  \BibitemOpen
  \href@noop {} {\bibinfo  {journal} {The linear approximation in $\Delta$ can
  be readily made in the Langeven equation for the field ladder operator where,
  at finite detuning, the parameter $G_\gamma$ acqures a complex factor. The
  absolute value of this factor differs from 1 by a corection of the order of
  $\Delta/\gamma$. Taking into acount this corection would result in the
  apearence of higher powers of $\Delta$ in the final stability condition, so
  that it is neglected, leading to the set of equations for the quadtatures,
  Eqs. (\ref{STAB3})}\ }\BibitemShut {NoStop}%
\end{thebibliography}

\end{document}